\documentclass[aps,amsmath,amssymb,preprintnumbers,nofootinbib,a4paper,prl,twocolumn]{revtex4-1}
\pdfoutput=1
\usepackage{amsthm}
\usepackage{graphicx}
\usepackage{color}
\usepackage{bbm}
\usepackage{pxfonts}
\usepackage{geometry}
\usepackage{epsfig}
\usepackage{multirow}

\usepackage[ margin=5pt, font=normalsize,labelfont=bf,justification=raggedright]{caption}

\definecolor{rossoCP3}{cmyk}{0,.88,.77,.40}

\begin{document}

\title{\Large  \color{rossoCP3}  Marginally Deformed Starobinsky Gravity} 
\author{Alessandro Codello}
\email{codello@cp3-origins.net} 
\author{Jakob Joergensen}
\email{joergensen@cp3-origins.net} 
\author{Francesco Sannino}
\email{sannino@cp3-origins.net} 
\author{Ole Svendsen}
\email{svendsen@cp3-origins.net} 

\affiliation{
\vspace{5mm} 
{ \color{rossoCP3}  \rm CP}$^{\color{rossoCP3} \bf 3}${\color{rossoCP3}\rm-Origins} \& the 
 {\color{rossoCP3} \rm Danish IAS} 
\mbox{ University of Southern Denmark, Campusvej 55, DK-5230 Odense M, Denmark}}
 \begin{abstract}
We show that quantum-induced marginal deformations of the Starobinsky gravitational action of the form $R^{2(1 -\alpha)}$, with $R$ the Ricci scalar and $\alpha$ a positive parameter smaller than one half, can generate sizable primordial tensor modes.
  We also suggest natural microscopic sources of these corrections and demonstrate that they generally lead to a nonzero and positive $\alpha$. Furthermore we argue, that within this framework, the scalar spectral index and tensor modes probe theories of grand unification including theories not testable at the electroweak scale. 
  \\~\\[.1cm]
{\footnotesize  \it Preprint: CP$^3$-Origins-2014-13 DNRF90 and DIAS-2014-13}
 \end{abstract}

\maketitle
 
%\section{Gravity Corrected Induced Inflation }
\label{nonminsec}

The fundamental origin of the inflationary paradigm is a central problem in cosmology \cite{Starobinsky:1979ty,Mukhanov:1981xt,Guth:1980zm,Linde:1981mu}. The simplest models of inflation typically introduce new scalar degrees of freedom. 
%, see  \cite{Martin:2013tda} for a review.  

An intriguing possibility is that gravity itself is directly responsible for the inflationary period of the universe. This requires one to go beyond the time-honored Einstein action, for example by adding a $R^2$-term as in the Starobinsky model \cite{Starobinsky:1979ty}. This approach is highly natural since it enables gravity itself to drive inflation without resorting to the introduction of new {\em ad hoc} scalar fields.
{The Starobinsky model in isolation predicts a nearly vanishing ratio of tensor to scalar modes ($r$), a result challenged by BICEP2 results \cite{Ade:2014xna}. However, independently on the validity of the BICEP2 result \cite{Flauger:2014qra}, 
it is of  fundamental importance to know the modifications on the Starobinsky model stemming from the embedding of a matter theory of particle physics in the gravitational theory.

According to \cite{Krauss:2013pha} cosmology can be used qualitatively to establish the quantization of gravity. In fact, by combining cosmological observations with an effective field theory (EFT) treatment of gravity \cite{Donoghue:1994dn,Burgess:2009ri} one can start estimating the parameters
entering gravity's effective action. An actual discovery of primordial tensor modes 
can therefore be used to determine these parameters at the inflationary scale, which may turn out to be close to the grand unification energy scale.

To lowest order, the effective action for gravity can be parametrized as
\begin{eqnarray}
\mathcal{S} =\int d^{4}x \sqrt{-g}&~&\left[-\frac{M_{p}^{2}}{2}R + h_0 R^2 + h_1 R^3 + \right.  \nonumber \\
&& \left. + c_0 C^2 + e_0 E + ... \right]
  \label{gravity_jordanaction} \,, \nonumber
\end{eqnarray}
where $M_p$ is the Planck mass. Beyond an expansion in the Ricci scalar $R$, we formally included the Weyl conformal tensor $C^2$ and the Euler four dimensional topological term $E$. However we can drop $E$ since it is a total derivative. Furthermore when gravity is quantized around the Friedmann Lemaitre Robertson Walker metric the Weyl terms are sub--leading since the geometry is conformally flat \cite{Iihoshi:2007uz}. We are left with an $f(R)$ form of the EFT. Higher powers of $R$, $C^2$ and $E$ are naturally suppressed by the Planck mass scale. If inflation occurs at energy scales much below the Planck scale the EFT is accurate. We must, however, take into account also marginal deformations including, for example, logarithmic corrections to the action above. Because of the similarity between the EFT description of gravity and the chiral Lagrangian for Quantum Chromo Dynamics we expect the quantum-induced logarithmic corrections to play a fundamental role for a coherent understanding of low energy gravitational dynamics at the inflationary scale. This is exactly what happens in hadronic processes involving pions at low energies.

We encode these ideas in a simple $f(R)$ form of the gravitational action formulated in the Jordan frame: 
\begin{eqnarray}
{\mathcal{S}_{J}} =\int d^{4}x \sqrt{-g} \left[-\frac{M_{p}^{2}}{2}R + h M_{p}^{4\alpha}\, R^{2(1-\alpha) } \right]
  \label{MDG} \,. \nonumber \\
\end{eqnarray}
We assume that $\alpha$ is a real parameter with $2 |\alpha |<1 $ and $h$ is a dimensionless parameter.
{We linearize the action via 
${\mathcal{S}_{J}} = \int d^{4}x \sqrt{-g} \left[f(y) + f^{\prime}(y) (R-y) \right]$ with 
$\displaystyle{f(R) = -\frac{M_{p}^{2}}{2}R + h M_{p}^{4\alpha}\, R^{2(1-\alpha) }}$. The equation of motion for $y$ implies $R=y$ provided $f^{\prime \prime} (y)$ does not vanish. Introducing the conformal mode $\psi = - f^{\prime}(y)$ with $V(\psi) = - y(\psi) \psi  - f(y(\psi))$ we arrive, after having introduced the mass dimension one real scalar field $\phi$ via $2\psi -  M^2_{P} = \xi \phi^2$ at: 
 \begin{eqnarray}
{\mathcal{S}_{J}} =\int d^{4}x \sqrt{-g} \left[-\frac{M_{p}^{2} + \xi \phi^2}{2}R - V(\phi) \right]
  \label{phi} \,, 
\end{eqnarray}
with $V(\phi) = \lambda \phi^4 \left(\frac{\phi}{M_p}\right)^{4\gamma}$, $\alpha = \gamma/(1+2\gamma)$
 and 
%\begin{equation}
$h^{1+2\gamma} =  \left(\frac{\xi}{4}\frac{1+2\gamma}{1+\gamma}\right)^{2(1+\gamma)} \frac{1}{\lambda(1+2\gamma)}$.}
%\,.
%\end{equation} }
 This provides the explicit relation between \eqref{MDG} and the effective quantum-corrected non-minimally coupled scalar field theory used in \cite{Joergensen:2014rya}. Here we can simply set the non-minimal coupling value of $\xi$ associated to the $\phi^2 R$ term to unity. This is valid since this action is equivalent to \eqref{MDG}, where only the two independent parameters $h$ and $\alpha$ appear. In this way $h$ depends only on $\gamma$ and $\lambda$. We will retain the explicit dependence on $\xi$ to ease the comparison with the results obtained in  \cite{Joergensen:2014rya}.

An important difference with respect to the non-minimally coupled theory of  \cite{Joergensen:2014rya} is the presence, already in the Jordan frame, of a kinetic term for $\phi$. In the $f(R)$ framework the kinetic term for $\phi$ is absent in the Jordan frame but it is, however, generated via the following conformal transformation of the metric:
\begin{align}
g_{\mu\nu}\rightarrow
%\tilde{g}_{\mu\nu}=
\Omega({\phi})^2 g_{\mu\nu},\quad\Omega({\phi})^2=1+\frac{\xi\phi^2}{M_{p}^2}\,\ . 
\end{align}
 This transformation allows us to rewrite both theories in terms of a propagating scalar field minimally coupled to ordinary Einstein gravity. This is the Einstein frame. Here the theory reads
\begin{widetext}
\begin{align}
\mathcal{S}_{E} &=\int d^{4}x \sqrt{-g}\left[-\frac{M^2_p}{2} R + \frac{1}{2} g^{\mu \nu} \partial_{\mu} \chi \partial_{\nu} \chi- U(\chi)  \right], \quad U(\chi) \equiv \left( \Omega^{-4}V \right) \left( \phi \left( \chi \right) \right)\,.
\label{einsteinframeaction} 
\end{align}
\end{widetext} 
The canonically normalized scalar field $\chi$ is related to $\phi$ via 
\begin{eqnarray}
\frac{1}{2} \left( \frac{d \chi}{d \phi} \right)^2 
%&= & \Omega^{-2} \left(\sigma + 3 M_p ^2 {\Omega'} ^2 \right) \nonumber \\
& = &\frac{M_p^2 \left(\sigma M_p^2 + \left(\sigma + 3 \xi \right) \xi \phi^2 \right)}{\left(M_p^2+\xi \phi^2 \right)^2}\,. \label{1defchi}
\end{eqnarray}
The map from the Jordan frame of $f(R)$ gravity to the Einstein frame with a canonically normalized field is obtained for $\sigma = 0 $ and agrees with other derivations \cite{Sotiriou:2008rp}. The case $\sigma =1$ corresponds to a $\phi$ with an initial kinetic term in the Jordan frame. Because of this $\sigma$ dependence it is, in principle, possible to disentangle the two original models.
To obtain an explicit relation between $\chi$ and $\phi$ we assume that inflation occurs at large values of the the scalar field, i.e. $\phi \gg \frac{M_p}{\sqrt{\xi}}$ yielding
\begin{equation}
\chi \simeq \kappa \, M_p \log \left( \frac{\sqrt{\xi} \phi }{M_p} \right)  \ , \quad \kappa = \sqrt{\frac{2\sigma}{\xi} +6} \, .
\end{equation}
For large values of the non-minimal coupling $\xi$ it is not possible to differentiate between the two values of $\sigma$. The Einstein frame potential takes the form
\begin{widetext}
\begin{align}
U \left( \chi\right)
&= \underbrace{ \frac{\lambda M_p^4}{\xi^2}\left(1+ \exp \left[\frac{ - 2 \chi}{\kappa M_p }\right] \right)^{-2}}_{\phi^4\text{-Inflation}} \underbrace{ \xi^{-2\gamma} \exp \left[ \frac{ 4 \gamma \chi}{\kappa  M_p}\right]}_{\text{Corrections from}\,\, \gamma}.
\end{align}
\end{widetext}
The underbraced '$\phi^4$-Inflation'-term refers to the potential one would obtain by setting $\gamma=0$, that is, non-minimally coupled $\phi^4$-Inflation. This limit corresponds to the Starobinsky model \cite{Starobinsky:1979ty}. 
 
The Starobinsky potential is recovered for $\alpha = 0$. For $ 0 <\alpha <0.5$ the potential grows exponentially and, as shown in \cite{Joergensen:2014rya}, it leads to a successful inflationary model with nonzero primordial tensor modes. On the other hand for $\alpha$ negative the potential is unable to produce enough e-folds. 
In Fig.~\ref{BICEP}  generic modifications of the Starobinsky model are confronted with BICEP2 and PLANCK data. We observe that cosmology may constrain the deformation parameter $\alpha$, and as we will show shortly, $\alpha$ holds information regarding the generic particle content embedded in this gravity model of inflation.
 \begin{figure}[t!]
\vskip .2cm
\begin{center}
\includegraphics[width=.40\textwidth]{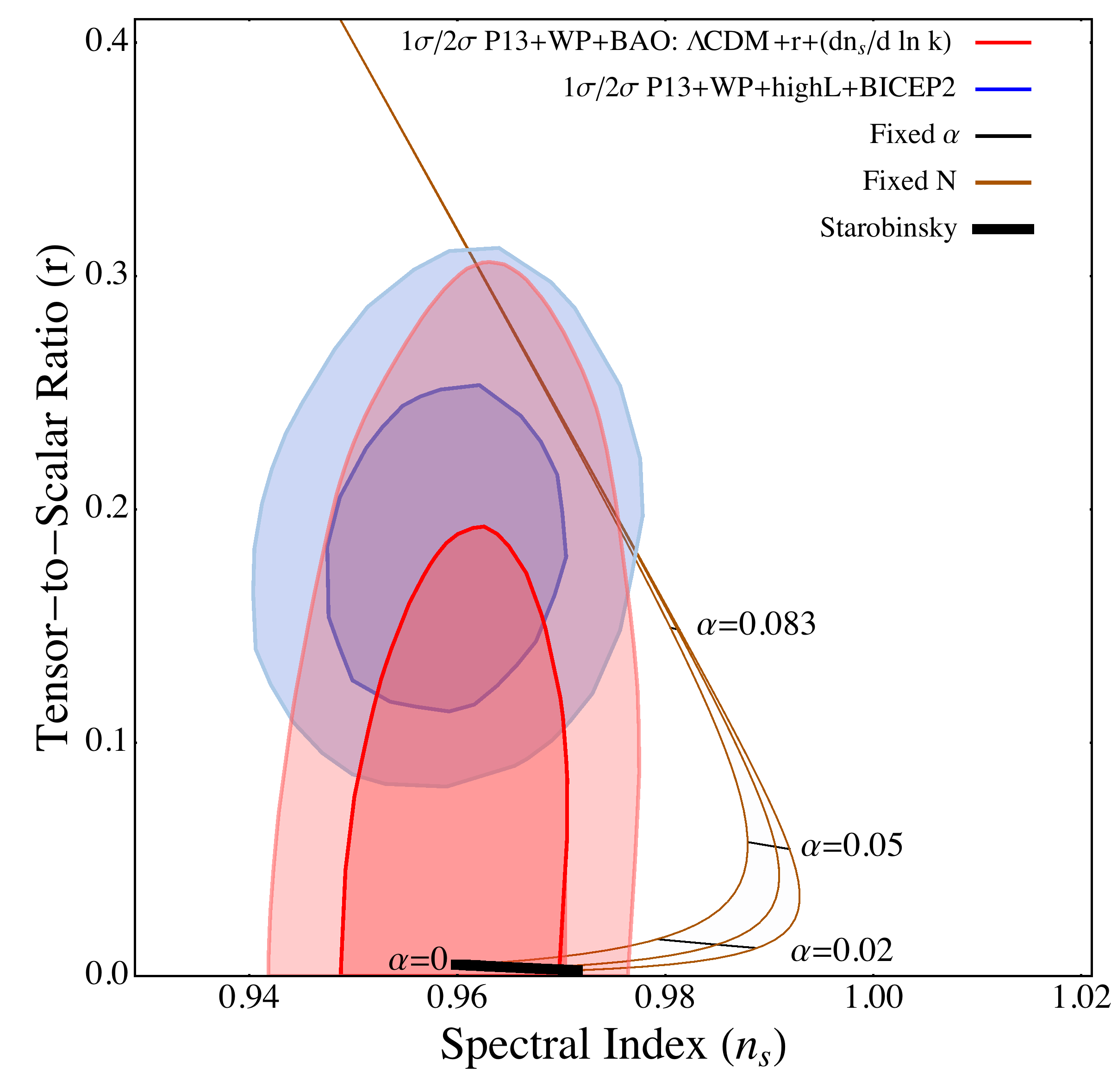}
\caption{Constraints on the parameter $\alpha$ by direct comparison with the combined fit of BICEP2 and Planck data provided in \cite{Ade:2014xna}.
%as well as the combined fit of BICEP2 and WMAP9 provided in \cite{Cheng:2014cja}. %See \cite{Audren:2014cea,Mortonson:2014bja} for caveats of the BICEP2 result.
} 
\label{BICEP}
\end{center}
\end{figure}

Different extensions of the original Starobinsky model including, for example, new degrees of freedom, have been investigated \cite{Ferrara:2014ima}. We believe that our approach is a minimal one which, as we shall see, also leads to relevant testable phenomenological implications. 

We now argue that these marginal deformations, needed from a purely phenomenological standpoint, arise naturally within a field-theoretical approach to quantum gravity. To gain insight we start by expanding \eqref{MDG} in powers of $\alpha$ and write
\begin{widetext}
\begin{eqnarray}
{\mathcal{S}_{J}} \simeq \int d^{4}x \sqrt{-g} \left\{-\frac{M_{p}^{2}}{2}R + h R^2\left[ 1 - 2 \alpha\log \left(\frac{R}{M_p^2} \right)\right] + {\cal O}(\alpha^2)\right\}
\label{MDGlog} \,. \nonumber \\
\end{eqnarray}
\end{widetext}
The logarithmic term is reminiscent of what one would obtain via  trace-log evaluations of quantum corrections.
There are several possible sources for these corrections. They may arise, for example, by integrating out matter fields, or they can arise directly from gravity loops.
To sum-up the entire series of logarithmic corrections, and hence recover the $R^{2(1-\alpha)}$, we expect that a renormalization group improved computation is needed.
This strongly suggests that one can determine $\alpha$ if a more fundamental theory were at our disposal.
In the absence of a full theory of quantum gravity we start here by comparing different predictions for the coefficient of the logarithmic term in \eqref{MDGlog} stemming out from: i) Integrating out minimally coupled non-interacting $N_S$ real scalar fields \cite{Codello:2011js} (only non--conformal invariant matter contributes); ii) gravity corrections via the effective field theory (EFT) approach \cite{Donoghue:1994dn,Burgess:2009ri,Donoghue:2012zc}; iii) gravity corrections within higher derivative gravity (HDG) \cite{Codello:2008vh}.
{The action \eqref{MDGlog} naturally arises after a direct computation of the quantum corrections stemming from the Tr $\log \Delta$ operator. This is deduced via heat kernel methods \cite{Avramidi}, where $\Delta$ is the laplacian arising as the Hessian of the minimally coupled scalar action, the Einstein-Hilbert action in the effective field theory case or the HDG action. The $R^2 \log R/\mu^2$ term emerges upon evaluating the operator on $S^4$. }
For dimensional reasons, further corrections  can be parametrized by an $h(R/\mu^2)\, R^2 $ term, where $h$ is now a function of $R/\mu^2$, with $\mu$ the renormalization scale.
{The explicit computations via heat kernel methods shows \cite{Avramidi} that} leading order quantum fluctuations indeed induce a logarithmic form for $h$ {as in (\ref{MDGlog})}.
{This fact alone immediately shows the link between the exponent $\alpha$ and the coefficient of the beta function of the coupling of the $R^2$ term, as a scale derivative with respect to the mass scale in (\ref{MDGlog}) shows.}
{But we can give a better argument noticing that,}
because $h$ depends on the ratio $R/\mu^2$, we have $2 R \partial_R h  = - \mu \partial_\mu h $
and %This permits
one can determine the $R$ dependence once the beta function, with respect to $\mu$, of $h$ is known.
{ Non--local $R^2 \log (-\square / \mu^2)$ quantum corrections can also be derived in a similar way \cite{Donoghue:1994dn}. { Cosmological applications of the non-local terms have recently been studied in \cite{Donoghue:2014yha}; other kinds of non-local actions, not generated by quantum loops, have also been considered in \cite{Deser:2007jk} and \cite{Maggiore:2014sia}. Here we are interested in the cosmological role of non-analytical terms which are present in the effective quantum gravity action.
Our analysis favors the idea that non-analytic terms are more significant for inflation than non-local ones.}
To the lowest order the beta function is $\mu \partial_\mu h = \frac{C}{(4\pi)^2}$ with $C$ a constant depending on the source of quantum corrections considered.
After an RG improvement, the equation for $h$ reads
\begin{equation}
R \,\partial_R h = - \frac{C}{2(4\pi)^2} h\,.
\label{RG}
\end{equation}
The improvement is related to the appearance of a factor $h(R/\mu^2)$ on the right hand-side of the equation above. If one sets $h(R/\mu^2) = 1$, on the right-hand side, we only obtain the first logarithmic correction of \eqref{MDGlog}.
Using \eqref{RG} we construct the log--resummed solution
\begin{equation}
h(R) = h(R_0) \left(\frac{R}{R_0}\right)^{-\frac{C}{2(4\pi)^2}} \,.
\label{a}
\end{equation}
Here $R_0 = \mu_0^2$ is a given renormalization scale. We therefore have $\alpha = \frac{C}{4(4\pi)^2}$ and the {\it constant} $h$ in \eqref{MDG} is $h(R_0)$.
If $C > 0$  this would naturally lead to a positive $ \alpha$. 
An explicit evaluation of $C$ gives \cite{Codello:2008vh,Codello:2011js}:
\begin{eqnarray}
C &=& \frac{N_S}{72} \qquad \textrm{minimally coupled scalars,} \nonumber\\
C &=& \frac{1}{4} \qquad\;\; \,\textrm{EFT gravity,} \label{cterms}\\
C &=& \frac{5}{36} \qquad\; \textrm{HDG}\nonumber.
\end{eqnarray}
Remarkably we deduce a {\it positive} exponent regardless of the underlying theory used to determine the associated quantum corrections to the gravitation action. { Massive particles (we consider scalars of mass $m$ for simplicity) lead to the beta function $\mu \partial_\mu h = \frac{C}{(4\pi)^2} (1+m^2 / \mu^2)^{-1}$ \cite{Codello:2008vh}. When the renormalisation scale is taken to be Planck's mass the effect of the mass term is negligible. Smaller renormalisation scales generally tend to reduce the value of $C$ and thus of $\alpha$, but in particular they don't affect it's sign.}

From \eqref{cterms} we deduce that quantum gravitational contributions can account, at most, for a 3\% increase in $r$ as compared to the original Starobinsky model. Therefore any larger value of $r$ can only be generated by adding matter corrections. This, in turn, can be used to constrain particle physics models minimally coupled to $f(R)$ gravity. Furthermore, as it is evident form Fig~\ref{BICEP}, for small $r$ the spectral index $(n_s)$ depends sensitively on the particular value of $\alpha$. We can therefore provide the following general constraint at the one sigma confidence level on the number of scalar fields minimally coupled to $f(R)$ gravity
\begin{equation}
  N_S \leq 85  \ .
  \label{nbound}
\end{equation}
The corresponding $r$ values cannot exceed $0.007$. To exemplify the power of our results we now compare \eqref{nbound} with popular models of grand unification (GUT) such as minimal SU(5) that features 34 scalars and (non)minimal SO(10) featuring (297) 109 scalars.  It is clear that only models with a low content of scalars are preferred by current experiments.  

Values of $r$ around and above $0.2$ can be achieved at two sigma confidence level only by allowing for the presence of thousands of scalars. This corresponds to the upper part of Fig.~\ref{BICEP}. Here one might hope to use non-minimal models of supersymmetric GUTs which would otherwise be physically excluded within the paradigm investigated here.

%
%GUT theories with more scalars may be considered, for example the supersymmetric SO(10) grand unified gauge theory discussed in \cite{Bajc:2005qe}. The number of scalar fields at the unification scale and above are of the order of $2(210_H + 2\times 126_H + 10_H + 16_{M}\times 3)$. Here the $H$ subscript labels the number of Higgs-like states needed to break the various gauge symmetries and $M$ labels the matter scalar fields. For this particular example we find $\alpha_{SUSY-SO(10)} \simeq 0.023$. 
%\\

{ We have pointed out that non-analytic terms are presents in the effective quantum action and can have a role in cosmology.} We have also checked that higher order terms of the type $R^3$ cannot lead to sizable nonzero tensor modes. Therefore we arrive at the general conclusion, that if inflation is driven by an $f(R)$ theory of gravity, a natural form for this function is the marginally deformed Starobinsky action provided in \eqref{MDG} with a positive $\alpha$ whose size is related to the microscopic theory dictating the trace-log quantum corrections. {This form can be tested by current and future experimental results and constitutes a natural generalization of the original Starobinsky action.  }     

 \noindent
%We thank Marco Nardecchia for useful discussions. 
Work supported by the Danish National Research Foundation DNRF:90 grant.

\end{document}